\documentclass[journal]{igtesymp}
\ifCLASSINFOpdf
\else
\fi
\usepackage[tight,footnotesize]{subfigure}
\usepackage{graphicx}

\usepackage[english]{babel}
\usepackage{lipsum}
\usepackage{amsmath, amssymb, amsthm}
\usepackage{color}

\usepackage{booktabs}	
\usepackage{siunitx}
\usepackage{tikz}		
\usepackage{pgfplots}
\usepackage{url}

\newtheorem{algorithm}{Algorithm}
\begin{document}
\title{Multi-objective free-form shape optimization of a synchronous reluctance machine}
\author{P.~Gangl\IEEEauthorrefmark{1},  S. K\"othe\IEEEauthorrefmark{1},
C.~Mellak\IEEEauthorrefmark{2}, A.~Cesarano\IEEEauthorrefmark{1}, and A.~M\"utze\IEEEauthorrefmark{2}
 \\ \vspace{0.3cm} \normalsize{
\IEEEauthorblockA{\IEEEauthorrefmark{1}Institute of Applied Mathematics, TU Graz, Steyrergasse 30, A-8010 Graz\\
 \IEEEauthorrefmark{2} Electric Drives and Machines Institute, TU Graz,
Inffeldgasse 18, A-8010 Graz\\
 E-mail: gangl@math.tugraz.at}}% <-this % stops a space
}



\IEEEaftertitletext{\vspace{-1cm}\noindent\begin{abstract} 
This paper deals with the design optimization of a synchronous reluctance machine to be used in an X-ray tube, where the goal is to maximize the torque, by means of gradient-based free-form shape optimization. The presented approach is based on the mathematical concept of shape derivatives and allows to obtain new motor designs without the need to introduce a geometric parametrization. We validate our results by comparing them to a parametric geometry optimization in \texttt{JMAG} by means of a stochastic optimization algorithm. While the obtained designs are of similar shape, the computational time used by the gradient-based algorithm is in the order of minutes, compared to several hours taken by the stochastic optimization algorithm. Finally, we show an extension of the free-form shape optimization algorithm to the case of multiple objective functions and illustrate a way to obtain an approximate Pareto front.
\end{abstract}
\noindent\vspace{\baselineskip}}

\maketitle
\thispagestyle{empty}\pagestyle{empty}

% \fi
\IEEEpeerreviewmaketitle

\section{Introduction}
In many industrial applications, the design of electric machines has to be tailored to the application at hand since off-the-shelf solutions are not available. The design of electric machines is usually based on engineering knowledge and is sometimes refined by geometric optimization. The most widely used approach is to introduce geometric parameters and optimize these, either using stochastic optimization algorithms or derivative-based methods, see \cite{BramerdorferOverview} for an overview article. While derivative-based optimization algorithms successively improve a given initial geometry by means of gradient information and are known to converge to a local optimum rather fast, stochastic algorihtms include random effects and are less prone to getting stuck in local optima. In practice, one is usually confronted with several conflicting objective functions thus making multiobjective optimization capabilities for finding a Pareto optimal set of designs important. The extension to a multiobjective setting is more straightforward in the case of many stochastic optimization algorithms, however it can also be achieved in the case of derivative-based methods \cite{doganay2019gradient}.

In recent years, non-parametric shape optimization methods based on the mathematical concept of shape derivatives \cite{DZ2} (often refered to free-form shape optimization approaches) have become a more and more popular tool for the design optimization of electric machines, see e.g. \cite{a_GALALAMEST_2015a, kuci2018phd, putek} for approaches using the finite element method or the recent work \cite{merkel2020shape} in the context of isogeometric analysis. In these approaches, the geometry is not parametrized by a finite number of scalar values, but the design variable is a set, e.g. the set of points occupied by ferromagnetic material in the rotor of an electric machine. Starting out from a given initial design, the design is updated by the action of a smooth vector field, thus allowing for any kind of design that is topologically equivalent to the initial design. This way, often new and innovative designs can be obtained. 

The purpose of this paper is two-fold: On the one hand, we extend the gradient-based multi-objective optimization method introduced in the case of a parametrized geometry in \cite{doganay2019gradient} to the case of free-form shape optimization. This allows to exploit the flexibility of free-form shape optimization methods, as well as their fast convergence properties also in the practically important case of multiple competing objective functions. 
On the other hand, we employ this method on both, the more standard single-objevtive case and, in the case of two objective functions, to find (Pareto-)optimal designs of a synchronous reluctance machine.
Comparing our results with the results obtained by a stochastic parameter optimization confirms the higher degree of flexibility and computational efficiency of our approach compared to parametric design optimization.

The rest of this paper is organized as follows: In Section~\ref{sec_probDesc} we introduce the problem at hand and state the mathematical model. We recall the main ingredients for a free-form shape optimization method and apply the algorithm to our problem in Section~\ref{sec_shapeOpti}. In Section~\ref{sec_multiobjShape} we show an extension of the gradient-based free-form shape optimization algorithm to the case of multiple objective functions before concluding in Section~\ref{sec_conclusion}.

\section{Problem description} \label{sec_probDesc}
\subsection{Physical model}

% \subsection{Application Example: Synchronous Reluctance Machine Operating Principle}
%*************%***************%***************%***************%***************
% A very good application of gradient-based free-form shape optimization is 
We consider the design optimization of a synchronous reluctance machine (SynRM), i.e., a motor that is based solely on the reluctance principle. This motor generates torque exclusively by a difference of reluctance between two axes, namely the \textit{d}-axis and the \textit{q}-axis (the location of the axes is defined by the number of poles of the machine). Thus, torque generation is not based on any transient behavior or quantity and a static magnetic field analysis is sufficient.
The machine under investigation is intended for the use in an X-Ray tube for medical applications.  The considered rotor will be operated in a vacuum and therefore must be built  of solid pieces of metal (as opposed to the commonly used steel sheet structure). Additionally, the air gap of the motor is unusually large (e.g., 10\,mm with an outer stator diameter of 130\,mm) decreasing the torque capability of the machine. Furthermore, the rotor has to withstand temperatures of up to $\SI{450}{\degreeCelsius}$.~\cite{MellakKrischanMuetze2018}

\begin{figure}
\centering
\includegraphics[width=0.8\columnwidth]{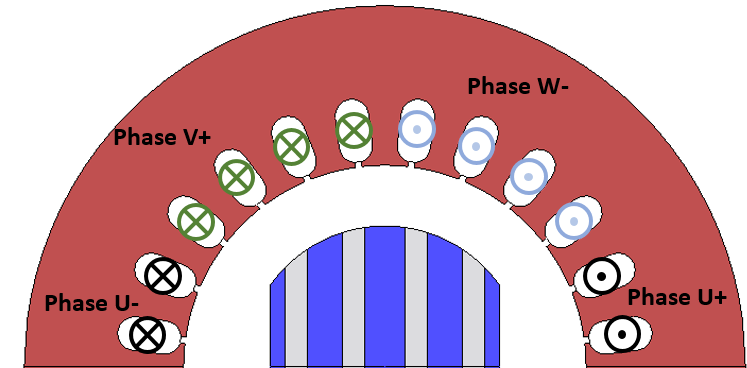}
    \caption{Upper half of synchronous reluctance machine with a three phase, two pole stator. The rotor consists of alternating magnetically conducting (blue) and non-conducting layers (gray).}
    \label{fig:stator}
\end{figure}

The synchronous reluctance machine is particularly suitable for such an application mainly due to its ruggedness and construction simplicity and the absence of rotor windings~\cite{xu1991}.
As per the operation mode of the machine quick acceleration and subsequent braking of a tungsten disk is required. Typically, this sequence takes at maximum 10\,s. Figure~\ref{fig:stator} shows the machine under investigation. The stator is a three phase stator with one pole pair, the rotor consists of alternating magnetically conducting (blue) and non-conducting layers (gray). The reference design parameters of the machine are stated in Table~\ref{tab:StatGeom}.

% \end{center}

\begin{table}
 \caption{Example case machine design parameters.}
\centering
\begin{tabular}{  l r  }
\toprule
Parameter & Value \\ 
\midrule
\textbf{Stator}  					& \\
  Inner radius 						& $\SI{26.5}{mm}$  \\
  Outer radius						& $\SI{47.5}{mm}$\\
  Number of slots 					& $24$  \\
  Number of phases 					& $3$  \\
  Number of poles 					& $2$  \\
  Axial length 						& $\SI{50}{mm}$   \\
  Winding type						& single-layer distributed  \\    
  No. of turns per slot 			& $64$  \\
  Phase resistance $R_{\text{S,\SI{20}{\degreeCelsius}}}$ & $\SI{7.1}{\ohm}$\\ 
  Rated voltage ${U}_{\text{eff}}$ 	& $\SI{230}{V}_\text{ac}$/ $\SI{400}{V}_\text{ac}$  \\
  Connection 						& star\\
 									& \\
 \textbf{Rotor} 					& \\
 Outer radius 						& $\SI{18.5}{mm}$   \\ 
 \bottomrule
\end{tabular}
    \label{tab:StatGeom}   
\end{table}

 \begin{figure}
\centering   
    \begin{tikzpicture}[>=latex,scale=0.80]
      \node[anchor=south] at (0,-1.84){\includegraphics[width=0.3\columnwidth]{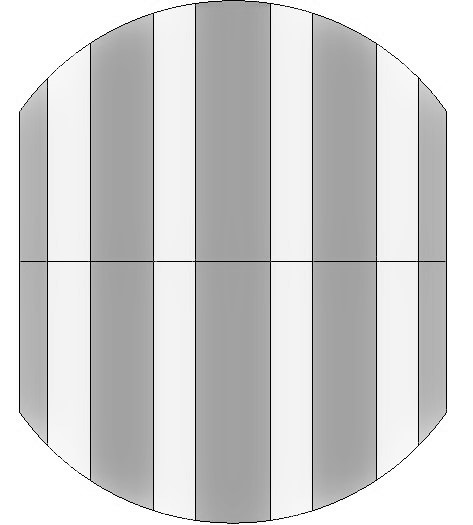}};
\draw[style=help lines] (0,0) (3,2);

\coordinate (vec_q) at (180:4.5); % q Achse
\coordinate (vec_d) at (90:4.5); % d Achse
\coordinate (vec_Fd) at (90:2.92403); % Flussverkettung_d
\coordinate (vec_Fq) at (100:3); % Flussverkettung_q
\coordinate (vec_Fdq) at (100:3); % Flussverkettung_dq
\coordinate (vec_Id) at (90:2.12132); % Strom_d
\coordinate (vec_Iq) at (180:2.12132); % Strom_q
\coordinate (vec_Idq) at (135:3); % Strom_dq
\coordinate (vec_Udq) at (210:4); % Spannung_dq

\draw[->,thick,gray] (0,0) -- (vec_q) node[below right] {q-axis};
\draw[->,thick,gray] (0,0) -- (vec_d) node[below right] {d-axis};
\draw[->,thick,black] (0,0) -- (vec_Fd) node [below right] {$\lambda_{\text{d}} = L_\text{d}\text{I}_\text{d}$};
\draw[->,thick,black] (0,2.92403) -- (vec_Fq) node [above right] {$\lambda_{\text{q}} = L_\text{q}\text{I}_\text{q}$};
\draw[->,thick,black] (0,0) -- (vec_Fdq) node [below left] {$\underline{\lambda}$};
\draw[->,thick,black] (0,0) -- (vec_Id) node [below right] {$\text{I}_{\text{d}}$};
\draw[->,thick,black] (0,0) -- (vec_Iq) node [above right] {$\text{I}_{\text{q}}$};
\draw[->,thick,black] (0,0) -- (vec_Idq) node [left] {$\underline{\text{I}}_{\text{s}}$};
% \draw[->,thick,black] (0,0) -- (vec_Udq) node [below right] {$\underline{U}_{\text{s}}$};
\draw [dashed,color=gray] (0,2.12132) -- (-2.12132,2.12132);
\draw [dashed,color=gray] (-2.12132,0) -- (-2.12132,2.12132);

% Winkel
\draw [red, thick] (0,1) arc [start angle=90, end angle=135, radius=1cm]
    node [midway, above, left] {$\beta$};

\end{tikzpicture}
    \caption{Vector diagram of a synchronous reluctance machine for the simplified model in d-q reference frame~\cite{binder2012}.}
    \label{fig:phasor_diagram}
  \end{figure}
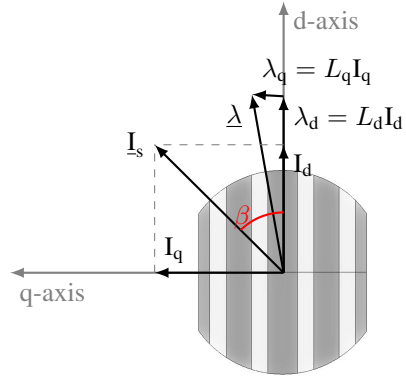

Figure~\ref{fig:phasor_diagram} shows the simplified vector diagram of a synchronous reluctance machine. The $d$-axis of the machine is the path with least reluctance, the $q$-axis is the path with the highest reluctance. In the \textit{d-q} axis theory, the torque is expressed as 
\begin{equation*}%\label{eq:synctorque_phiI} 
T = \frac{3N_p}{2}(\lambda_{\text{d}}{\text{I}_{\text{q}}} - \lambda_{\text{q}}{\text{I}_{\text{d}}}),
\end{equation*}
where $N_p$ denotes the number of pole pairs, $\lambda_{\text{d}}$ and $\lambda_{\text{q}}$ are the magnetic flux linkages, and $\text{I}_{\text{d}}$ and $\text{I}_{\text{q}}$ are the currents in $d$-axis and $q$-axis direction, respectively. Alternatively, using the inductances $L_{\text{d}}$ and $L_{\text{q}}$ as well as the stator current $\text{I}_{\text{s}}$ and current angle $\beta$, the torque is expressed as

\begin{equation}\label{eq:synctorque} 
T = \frac{3N_p}{4}(L_{\text{d}} - L_{\text{q}})\text{I}_{\text{s}}^2\text{sin}(2\beta)\,.
\end{equation}

Evidently, as per~\eqref{eq:synctorque}, assuming linear lossless behaviour and a fixed stator current $\text{I}_{\text{s}}$ the maximum torque can be achieved with a machine current angle~$\beta$ (angle between current vector and d-axis of the machine, Fig.~\ref{fig:phasor_diagram}) of \SI{45}{\degree}.~\cite{spargo2013}

\subsection{Optimization goal}
A static analysis is chosen to calculate the reluctance torque. Therefore, a current is impressed on the windings according to Table~\ref{tab:currentSheet}. Subsequently, the rotor is rotated and fixed clockwise to create the optimal current angle $\beta$ of \SI{45}{\degree}. The objective is to increase the torque with the given stator at a constant current and air gap length at the optimum current angle $\beta$.
The number of conducting and non-conducting layers remains unchanged. Solely the shape of each individual layer is subject to the optimization as to increase the $d$-axis inductance $L_{\text{d}}$ while, ideally, decreasing the $q$-axis inductance $L_{\text{q}}$ at the same time.

\begin{table}[!ht]
 \caption{The current values for each winding.}
\centering
\begin{tabular}{c c c}
\toprule
U-Phase & V-Phase & W-Phase  \\ 
\midrule
  12\,A 	&  -6\,A & -6\,A  \\ 
 \bottomrule
\end{tabular}
    \label{tab:currentSheet} 
\end{table}

\subsection{Mathematical model}
We consider a two-dimensional cross-section of the machine in the setting of 2D magnetostatics, i.e., $\mathbf B = \mbox{curl} \mathbf A$ where the magnetic vector potential is of the form $\mathbf A = (0, 0, u(x_1, x_2))^\top$. Let $D \subset \mathbb R^2$ denote the computational domain which comprises the two-dimensional cross section of the machine as well as a surrounding air region, and let $\Omega \subset D$ denote the ferromagnetic parts of the machine. The mathematical design optimization problem reads
\begin{align} \label{eq_opti_cost}
    \underset{\Omega \in \mathcal A}{\mbox{max  }} T(u) 
\end{align}
\begin{align}
    \begin{aligned} \label{eq_opti_pde}
    \mbox{s.t. } -\mbox{div}( \nu_\Omega(x, |\nabla u|)\nabla u) &= J_{i}, \quad x \in D, \\
                            u &= 0, \; \quad x \in \partial D,
    \end{aligned}
\end{align}
where $T$ represents the torque for the considered rotor position, $\mathcal A$ is a set of admissible shapes, $J_i$ represents the impressed current density and the magnetic reluctivity is defined piecewise as 
\begin{align*}
    \nu_\Omega(x,s) = \begin{cases}
                        \hat{\nu}(s) & x \in \Omega, \\
                        \nu_0 & x \in D \setminus \overline \Omega.
                      \end{cases}
\end{align*}
Here, $\hat{\nu}$ is a nonlinear function which represents the magnetic reluctivity of the ferromagnetic material, and $\nu_0$ corresponds to the magnetic reluctivity of air. The partial differential equation (PDE) constraint \eqref{eq_opti_pde} admits a unique solution under natural assumptions on the nonlinear function $\hat{\nu}$ \cite{PechsteinJuettler2006}. Note that the torque $T$ depends on the shape $\Omega$ of the ferromagnetic components via the solution to the PDE constraint \eqref{eq_opti_pde}. Denoting the unique solution to \eqref{eq_opti_pde} for given $\Omega \in \mathcal A$ by $u_\Omega$, we define the reduced cost function $\mathcal T(\Omega) := T(u_\Omega)$.

\section{Free-form shape optimization} \label{sec_shapeOpti}
We propose a free-form shape optimization algorithm based on the mathematical concept of shape derivatives, which is capable of improving the shape of a given initial geometry without the need of defining geometric parameters. We will outline the main ingredients to the method in the following. We introduce the theory for a general cost function $\mathcal J$ and will choose $\mathcal J := - \mathcal T$ later in Section~\ref{sec_numerics_1func}.

\subsection{Shape derivative} \label{sec_shapeDer}
The shape derivative of a general shape function $\mathcal J = \mathcal J(\Omega)$ represents the sensitivity of $\mathcal J$ when the domain $\Omega$ is perturbed by the action of a given vector field $V$. Given a smooth vector field $V$ which is defined on $D$, let $\Omega_t = (\mbox{id} +tV)(\Omega)$ denote the perturbed domain for $t > 0$. The shape derivative of $\mathcal J$ in the direction given by $V$ is defined as
\begin{align} \label{def_shape_der}
    d \mathcal J(\Omega; V) :=  \underset{t \searrow 0}{\mbox{lim }} \frac{\mathcal J(\Omega_t) - \mathcal J(\Omega)}{t},
\end{align}
provided that this limit exists and the mapping $V \mapsto d\mathcal J(\Omega; V)$ is linear and continuous \cite{DZ2}.

The shape derivative for problem \eqref{eq_opti_cost}--\eqref{eq_opti_pde} can be derived in an analogous way as it was done in \cite{a_GALALAMEST_2015a} and, for a vector field $V$ that is only supported on the rotor, reads
\begin{align}
    \begin{aligned} \label{eq_shapeDerFormula}
    d& \mathcal J(\Omega; V) =\\ %- \int_{\textcolor{red}{D}} (J_i \mbox{div}(V) \textcolor{red}{+ \nabla J_i \cdot V }) p \, \mbox dx \\
				&+\int_{D} \nu_\Omega(x,|\nabla u|) \left( (\mbox{div}\,V) I -\partial V^T-\partial V \right)\nabla u\cdot \nabla p \,\mbox dx\\
				& -\int_{D} \frac{\partial_s \nu_\Omega(x,|\nabla u|)}{|\nabla u|} (\partial V^T\nabla u \cdot \nabla u) (\nabla u\cdot \nabla p)\, \mbox dx.
    \end{aligned}
\end{align}
Here, $p$ denotes the solution to the adjoint equation which for the case of the maximization of the torque reads in its strong form 
\begin{align}
    \begin{aligned} \label{eq_adjoint}
    \mbox{s.t. } -\mbox{div}\left(A_{\Omega}(u) \nabla p  \right) &= \frac{\partial T}{\partial u}, \quad x \in D, \\
        p &= 0, \; \quad \; \; \, x \in \partial D. 
    \end{aligned}
\end{align}

with 
\begin{align*}
    A_{\Omega}(u):=  \nu_\Omega(x, |\nabla u|) I+\frac{\nu_\Omega'(x,|\nabla u|)}{|\nabla u|}\nabla u \otimes \nabla u.
\end{align*}

\subsection{Descent direction} \label{sec_descDir}
Given a closed formula for the shape derivative, a descent vector field $V$ can be obtained by solving an auxiliary boundary value problem as follows. Let $X$ be a Hilbert space and $b : X \times X \rightarrow \mathbb R$ a symmetric and positive definite bilinear form. Then the solution $W \in X$ to the variational problem 
\begin{align} \label{eq_auxbvp}
    b(W, V ) = - d \mathcal J(\Omega; V) \; \forall V \in X
\end{align}
is a descent direction since it satisfies by construction
\begin{align*}
    d \mathcal J(\Omega; W) = - b(W, W)  < 0.
\end{align*}
Thus, it follows from the definition in \eqref{def_shape_der} that perturbing $\Omega$ a small distance into the direction $W$ will yield a decrease of the cost function $\mathcal J$. 

The user has some degrees of freedom in the choice of the bilinear form $b(\cdot, \cdot)$ as well as the space $X$. Common choices include $X = H^1(D,\mathbb R^2)$ and $b(W,V) = \int_D \partial W : \partial V + W \cdot V \mbox dx$ or $b(W, V) = \int_D C \epsilon(W) : \epsilon(V) + W \cdot V \mbox dx$ where $\epsilon(V) = \frac{1}{2} (\partial V + \partial V^\top)$ and $C$ is a fourth-order elasticity tensor. The latter choice is known to preserve mesh quality better compared to other choices of $b(\cdot, \cdot)$ \cite{a_IGSTWE_2018a}. 
An alternative strategy for extracting a descent direction which also allows for the extension to multiple objective functions will be discussed in Section~\ref{sec_multiobj_descent}.

\subsection{Numerical results} \label{sec_numerics_1func}
The procedure outlined in Sections \ref{sec_shapeDer} and \ref{sec_descDir} constitutes the following free-form shape optimization algorithm for minimization of shape function $\mathcal J = \mathcal J(\Omega)$:
\begin{algorithm} \label{algo_singleobj} Given initial design $\Omega_0$, cost function $\mathcal J$, tolerance $tol$, $k=0$.
\begin{enumerate}
 \item Solve state equation \eqref{eq_opti_pde} and adjoint equation \eqref{eq_adjoint}
 \item Compute shape derivative $d\mathcal J(\Omega_k; V)$ given in \eqref{eq_shapeDerFormula}
 \item Compute shape gradient $W$ as solution to \eqref{eq_auxbvp}
 \item If $\|W\| < tol$ then stop \\
 else set $\Omega_{k+1} = (\mbox{id}+t W)(\Omega_k)$ where $t = \mbox{max}\{1, \frac{1}{2}, \frac{1}{4}, \frac{1}{8}, \dots \}$ such that $\mathcal J(\Omega_{k+1}) < \mathcal J(\Omega_k)$.
 \item $k \leftarrow k+1$ and go back to 1)
\end{enumerate}
\end{algorithm}
In step 4) the parameter $t$ is chosen by a line search in order to guarantee a descent of the cost function $\mathcal J$.

We applied Algorithm \ref{algo_singleobj} to problem \eqref{eq_opti_cost}--\eqref{eq_opti_pde}, i.e. we chose to minimize $\mathcal J(\Omega) := - \mathcal T(\Omega)$, using the finite element software package \texttt{NGSolve} \cite{Schoeberl2014}. In particular, we used the automated shape differentiation capabilities provided by \texttt{NGSolve} which enables the automated computation of the shape derivative $d \mathcal J(\Omega; V)$ for a large class of PDE-constrained shape optimization problems~\cite{GanglSturmNeunteufelSchoeberl2020}.

\begin{figure}
\begin{center}
    \begin{tabular}{c}
        \includegraphics[width = .45\textwidth, trim = 0 0 0 0, clip]{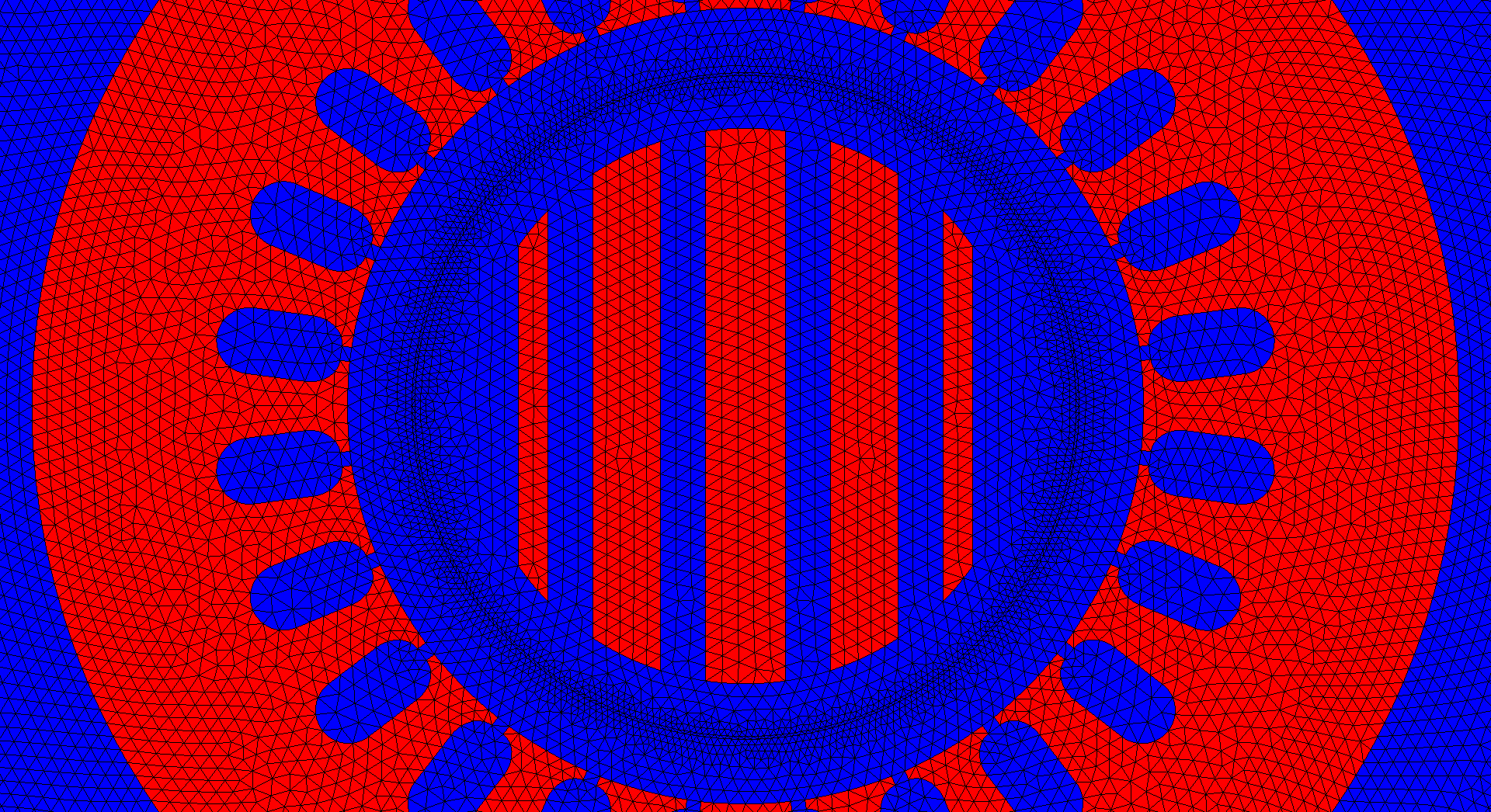} \vspace{2mm} \\ \includegraphics[width = .45\textwidth, trim = 0 0 0 0, clip]{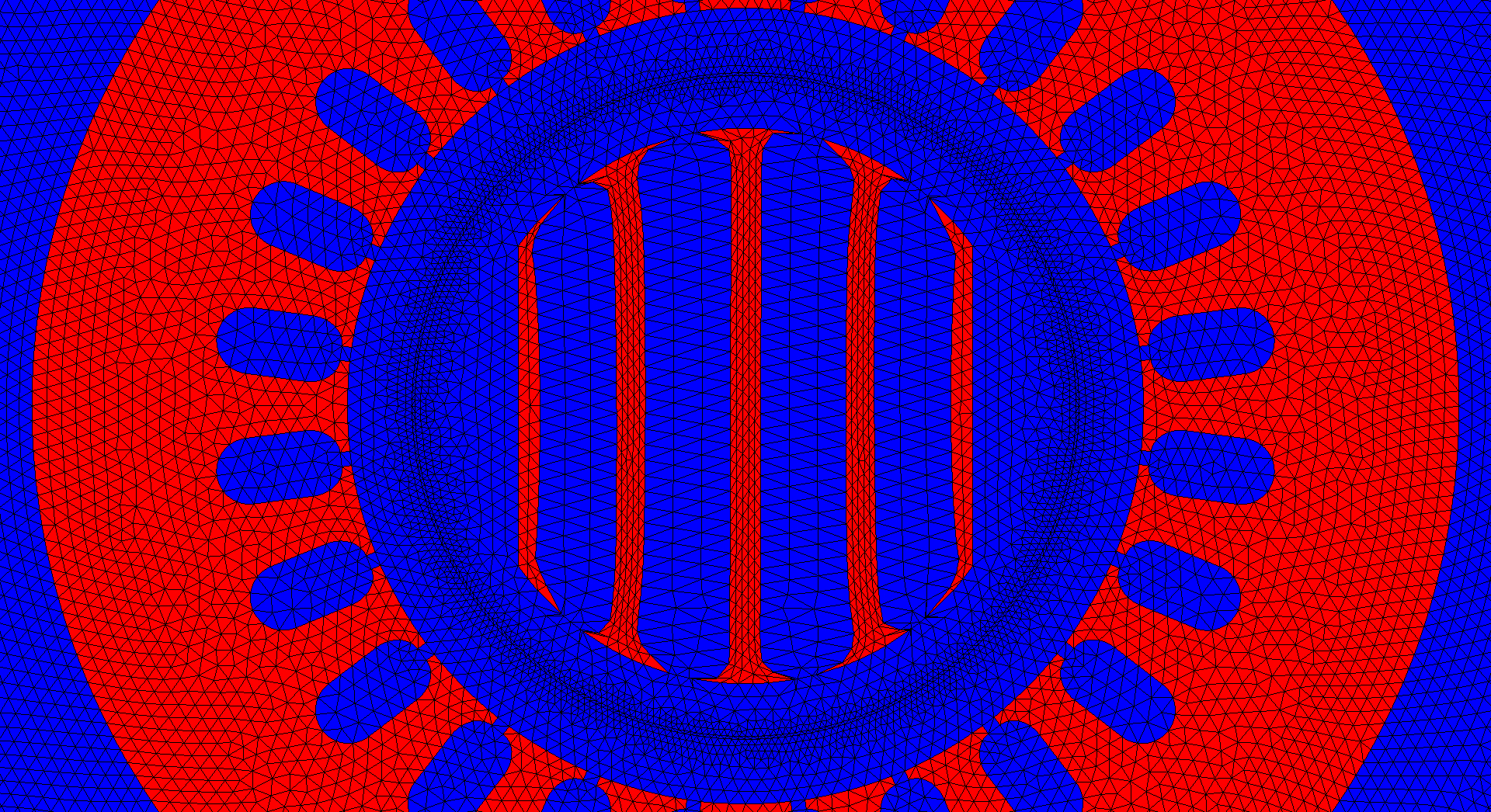}
    \end{tabular}
\end{center}
    \caption{Top: Initial design of rotor, $T = 1.007$~Nm. Bottom: Optimized design obtained after 70 iterations of Algorithm \ref{algo_singleobj}, $T = 1.270$~Nm.}
    \label{fig_designSingleObj}
\end{figure}

For the space $X$ in \eqref{eq_auxbvp}, we chose the space of all vector-valued $H^1$ functions defined on the rotor of the machine whose normal component vanishes on the top and bottom boundary parts of the rotor and which vanish at the left and right boundary parts. For the bilinear form $b(\cdot, \cdot)$ we chose the $H^1$ inner product
\begin{align*}
b(V,W):&= \int_{D_{rot}} \partial V : \partial W  + \frac{1}{100} V\cdot W\;\mbox dx,
\end{align*}
 where $D_{rot}$ denotes the union of the five iron and four air layers as depicted in Figure \ref{fig_designSingleObj}. 
The results obtained after 70 iterations of Algorithm \ref{algo_singleobj} are depicted in Figure \ref{fig_designSingleObj}. The torque was increased by about 26$\%$ from 1.007~Nm to 1.270~Nm. The computational time to obtain the optimized design was about 10 minutes on a single core.

\subsection{Validation} \label{sec_validation}
\begin{figure}
\centering   
\includegraphics[width=.3\textwidth]{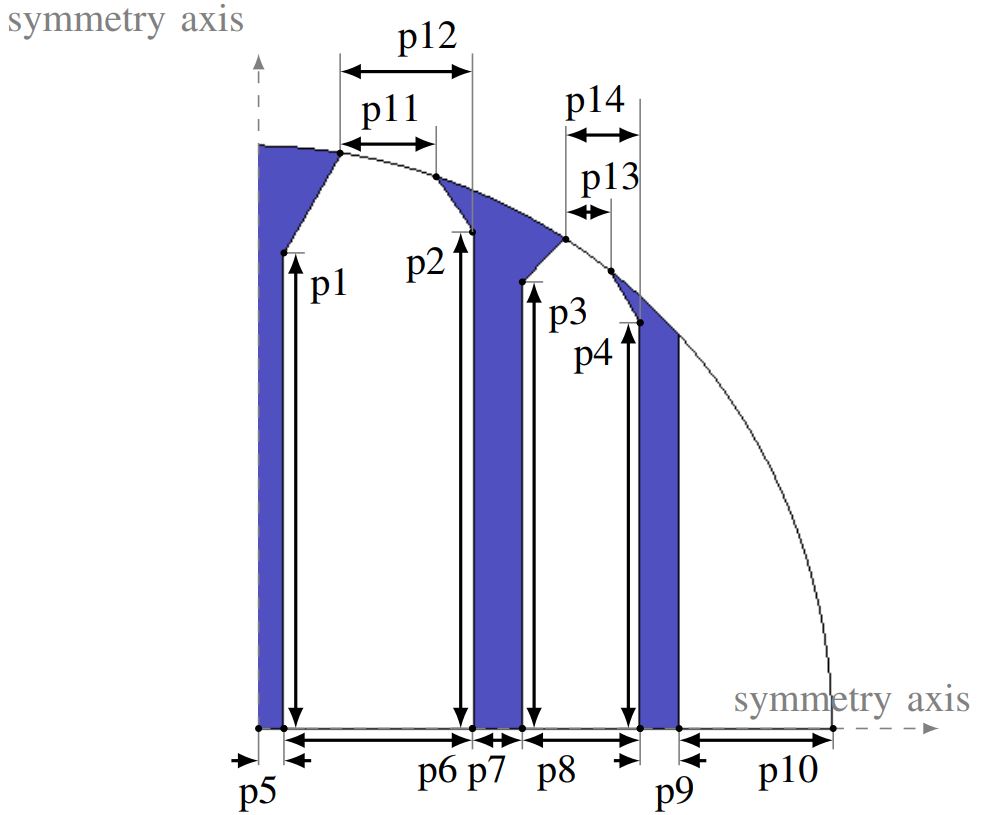}
    \caption{Geometric parameters used for genetic algorithm.}
    \label{fig_geoPara}
\end{figure}

\begin{figure}
    \begin{tabular}{cc}
        \includegraphics[width=.22\textwidth]{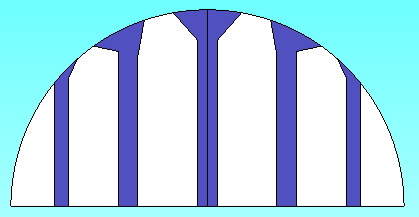}& \includegraphics[width=.22\textwidth]{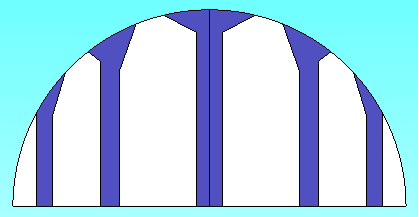} \\
        (a) & (b)\\
        \includegraphics[width=.22\textwidth]{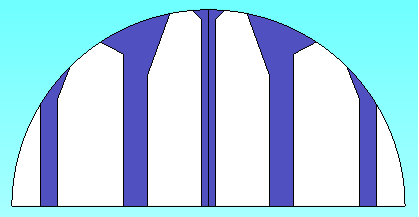}& \includegraphics[width=.22\textwidth]{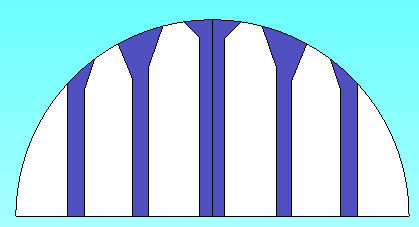} \\
        (c) & (d)\\
    \end{tabular}
    \caption{Best results obtained by genetic algorithm in \texttt{JMAG} based on geometric parametrization of Fig.~\ref{fig_geoPara} after 300 generations. (a) best design, $T=1.2119$Nm. (b) second best design, $T=1.2091$Nm. (c) third best design, $T=1.2082$Nm. (d) fourth best design, $T=1.2067$Nm.}
    \label{fig_optiDesignsCM}
\end{figure}

In this section, we validate the results obtained in our numerical experiments by comparing them to an optimization run in \texttt{JMAG} \cite{JMAG}. Motivated by the results of the gradient-based optimization, see Fig.~\ref{fig_designSingleObj}, we parametrized our rotor geometry by means of 14 geometric parameters under symmetry conditions, see Fig.~\ref{fig_geoPara}, and ran a genetic algorithm which is built into \texttt{JMAG} to maximize the torque. We started with a population size of 300 and ran the algorithm for 50 generations, allowing for 60 children in each generation. 
The computational time used by the genetic algorithm was about 19 hours and a total of 15000 designs were examined. The four designs with the highest torque values are depicted in Fig.~\ref{fig_optiDesignsCM}.
It can be seen that the best designs are similar to the design we obtained by the gradient-based algorithm (Fig.~\ref{fig_designSingleObj}), but also that the torque values were not quite reached. While one might be tempted to explain such a discrepancy by the fact that different simulation tools were used, we mention that the calculated torques in the two simulation softwares (\texttt{NGSolve} and \texttt{JMAG}) showed a good match for the initial geometry. Thus, it seems like the design in Fig.~\ref{fig_designSingleObj} is superior to those obtained by the genetic algorithm in \texttt{JMAG} since more general geometries can be obtained. Of course, the computation time of 19 hours could be reduced by reducing the parameters of the genetic algorithm, however the general order of magnitude remains. Finally note that, since the choice of the geometric parameters was inspired by Fig.~\ref{fig_designSingleObj}, the designs in Fig.~\ref{fig_optiDesignsCM} would have been unlikely to be found without the knowledge provided by the free-form shape optimization algorithm.

\section{Multi-objective shape optimization} \label{sec_multiobjShape}
In this section we consider an extension of the gradient-based free-form shape optimization method presented in Section~\ref{sec_shapeOpti} to the setting of multiple objective functions. We show how to compute a descent vector field $W$ that assures a descent with respect to several objective functions and use this approach in order to obtain an approximation of the Pareto front. We apply the method to the bi-objective free-form shape optimization problem
\begin{align*}
    \underset{\Omega}{\mbox{min }} \left( \begin{array}{cc} \mathcal J_1(\Omega) \\ \mathcal J_2(\Omega)  \end{array} \right)
\end{align*}
where $\mathcal J_1(\Omega) := -\mathcal T(\Omega)$ corresponds to the negative of the torque related to $\Omega$ and $\mathcal J_2(\Omega) := \mbox{Vol}(\Omega)$ denotes the volume of the ferromagnetic subdomains of the machine.

\subsection{Multi-objective descent direction} \label{sec_multiobj_descent}
Given two shape functions $\mathcal J_1$, $\mathcal J_2$ and their corresponding shape derivatives $d \mathcal J_i(\Omega; V)$, $i=1,2$, we want to find a vector field $W$ such that
\begin{align*}
    d \mathcal J_1(\Omega; W)<0 \quad \mbox{and} \quad d \mathcal J_2(\Omega; W)<0.
\end{align*}
We extend the ideas introduced in the framework of parametric shape optimization in \cite{doganay2019gradient} to the setting of free-form shape optimization. For that purpose, we consider a finite element discretization using piecewise linear and globally continuous finite elements on a triangular mesh. Denoting the corresponding hat basis functions by $\varphi_1, \dots, \varphi_n$ where $n$ is the number of mesh points and $\Phi_i = (\varphi_i,0)^\top$, and $\Phi_{n+i} = (0, \varphi_i)^\top$, $i = 1, \dots, n$, we have that 
\begin{align*}
    \{ \Phi_1, \dots,  \Phi_{2n} \}
\end{align*}
is a basis for the set of all two-dimensional vector fields on the mesh. Thus, after discretization each vector field $W_h$ can be written as $W_h = \sum_{i=1}^{2n} W_i \Phi_i$ with the coefficient vector $\underline W:=(W_1, \dots W_{2n})^\top$. Note that we can identify the finite element function $W_h$ with its coefficient vector $\underline W$. In order to obtain a discrete bi-descent direction $W_h$, we solve the auxiliary optimization problem to find $(\rho, \underline W) \in \mathbb R \times \mathbb R^{2n}$ 
\begin{align} \label{eq_aux_opti}
    \begin{aligned}
    \underset{\rho, \underline W}{\mbox{min}} \; &\rho + \frac{1}{2} \sum_{i=1}^{2n} W_i^2, \\
    \mbox{s.t.} \qquad & d \mathcal J_1(\Omega;W_h) \leq \rho,\\
    & d \mathcal J_2(\Omega;W_h) \leq \rho.
    \end{aligned}
\end{align}
Due to the linearity of the shape derivatives $d \mathcal J_i(\Omega; W_h)$ with respect to $W_h$, the solution $(\rho, \underline W) = (0, \mathbf 0) \in \mathbb R \times \mathbb R^{2n}$ is a feasible point of \eqref{eq_aux_opti}. Therefore, it follows that the solution $(\rho, \underline W)$ to \eqref{eq_aux_opti} satisfies $d \mathcal J_i(\Omega; W_h) \leq \rho \leq 0$, $i =1, 2$, thus giving a bi-descent direction $W_h$ whenever the optimal $\rho$ is negative. The second term in the cost function of \eqref{eq_aux_opti} is meant to keep the norm of $W_h$ bounded.

We remark that, in contrast to the widely used weighted-sum method, this approach is also feasible for finding non-convex parts of a Pareto front \cite{doganay2019gradient}.
Of course, an extension of this approach to account for more than two cost functions $\mathcal J_1, \dots, \mathcal J_N$ is straightforward.

\subsection{Obtaining a Pareto front}
Proceeding as described in Section~\ref{sec_multiobj_descent} allows to obtain a bi-descent direction $W_h$. Thus, starting out from an initial design, iteratively computing a bi-descent vector field and moving the interface a small distance in the direction given by this vector field constitutes a gradient-based free-form shape optimization algorithm for two cost functions. When no further decrease can be obtained, a Pareto optimal point is found.

In order to obtain many Pareto optimal points, one could start with many different initial designs. However, it turns out to be more convenient to proceed as follows: Consider different scalings of the two objective functions, i.e. apply the gradient-based biobjective descent algorithm for the two objective functions $\mathcal J_1$ and $w \mathcal J_2$ with different values of the weight $w$, see also \cite{doganay2019gradient}. Each choice of the weight $w$ corresponds to a run of the bi-objective descent algorithm and will yield a point on the Pareto front.

\subsection{Numerical results}
The proposed algorithm to obtain an approximation of a Pareto front consists in a loop over different weights $w$ where each iteration uses an algorithm similar to Algorithm \ref{algo_singleobj} to obtain an optimized design. In contrast to Algorithm \ref{algo_singleobj}, however, here the descent direction is obtained by solving the auxiliary optimization problem \eqref{eq_aux_opti} rather than an auxiliary boundary value problem of the form \eqref{eq_auxbvp}. The algorithm reads as follows:
\begin{algorithm} \label{algo_multiobj} Given initial design $\Omega_0$, cost functions $\mathcal J_1, \mathcal J_2$, tolerance $tol$, set of weights $\{w_1, \dots, w_M\}$.\\
For $j = 1, \dots, M$:
\begin{enumerate}
 \item If $j>M$ then stop \\
 else set $\tilde{\mathcal J_1} \leftarrow \mathcal J_1$, $ \tilde{\mathcal J_2} \leftarrow w_j \mathcal J_2$.
 \item Set $k \leftarrow  0$, $\Omega_0^{(j)} \leftarrow  \Omega_0$
 \item For $k = 0, 1, 2, \dots$
    \begin{enumerate}
    \item[(i)] Solve state equation \eqref{eq_opti_pde} and adjoint equation~\eqref{eq_adjoint}
    \item[(ii)] Compute shape derivatives $d \tilde{\mathcal J_1}(\Omega_k^{(j)}; V)$, $d \tilde{\mathcal J_2}(\Omega_k^{(j)}; V)$
    \item[(iii)] Compute bi-objective descent direction $W_h$ as solution to \eqref{eq_aux_opti} with $d \tilde{\mathcal J_1}(\Omega_k^{(j)}; \cdot)$, $d \tilde{\mathcal J_2}(\Omega_k^{(j)}; \cdot)$
    \item[(iv)] If $\|W_h\| < tol$ then $j \leftarrow j+1$ and go to 1) \\
    else set $\Omega_{k+1}^{(j)} = (\mbox{id}+t W_h)(\Omega_k^{(j)})$ where $t = \mbox{max}\{1, \frac{1}{2}, \frac{1}{4}, \frac{1}{8}, \dots \}$ such that $\mathcal J_i(\Omega_{k+1}^{(j)}) < \mathcal J_i(\Omega_k^{(j)})$, $i=1,2$.
    \end{enumerate}
\end{enumerate}

\end{algorithm}

In our implementation, we solved the quadratic optimization problem involving linear inequality constraints \eqref{eq_aux_opti} by means of a sequential least squares programming optimization algorithm using the functionality \texttt{scipy.optimize(...)}. In order to reduce computation time, we restricted problem \eqref{eq_aux_opti} to the degrees of freedom on the material interfaces which are subject to optimization and neglected the interior degrees of freedom. This is motivated by the fact that a movement of points inside a subdomain does not alter the shape. Proceeding like this, we obtain a deformation vector field that is only supported on the material interfaces and vanishes on all interior mesh nodes. In order to avoid intersection of the mesh when updating the geometry, we extend the vector field from the interfaces to the whole rotor domain by harmonic extension, i.e., by solving an elliptic PDE. As additional constraints, we imposed the linear equality constraints that the normal component of the vector field on the boundary on the rotor domain vanishes, i.e. $W_x(z)n_x(z) + W_y(z)n_y(z) = 0$ for all mesh points $z \in \partial D_{rot}$. These constraints ensure that the radius of the rotor remains unchanged. 

\begin{figure}
\begin{tabular}{cc}
    \includegraphics[width=0.22\textwidth, trim = 10 0 30 0, clip]{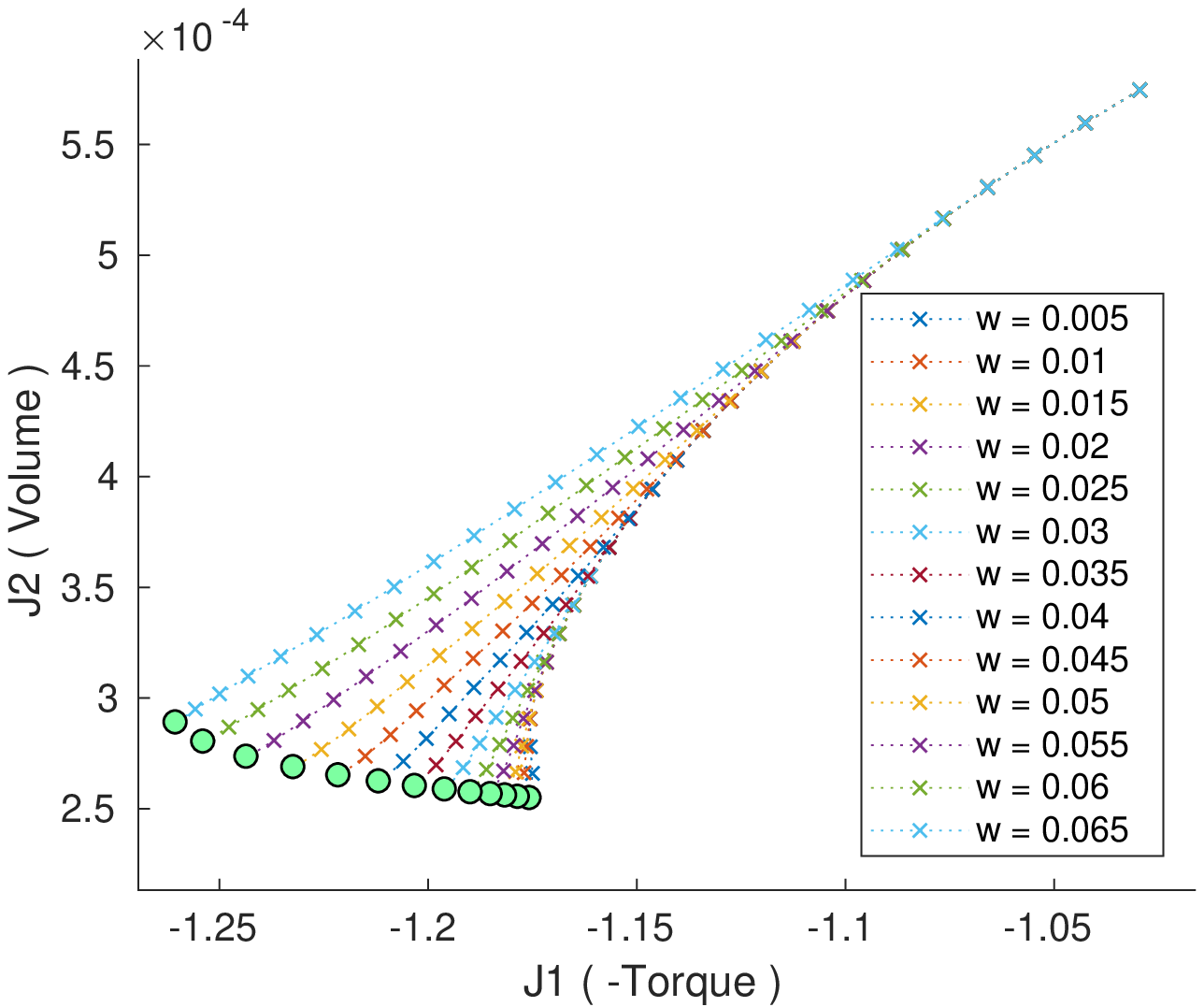}&
    \includegraphics[width=0.23\textwidth, trim = 0 0 20 0, clip]{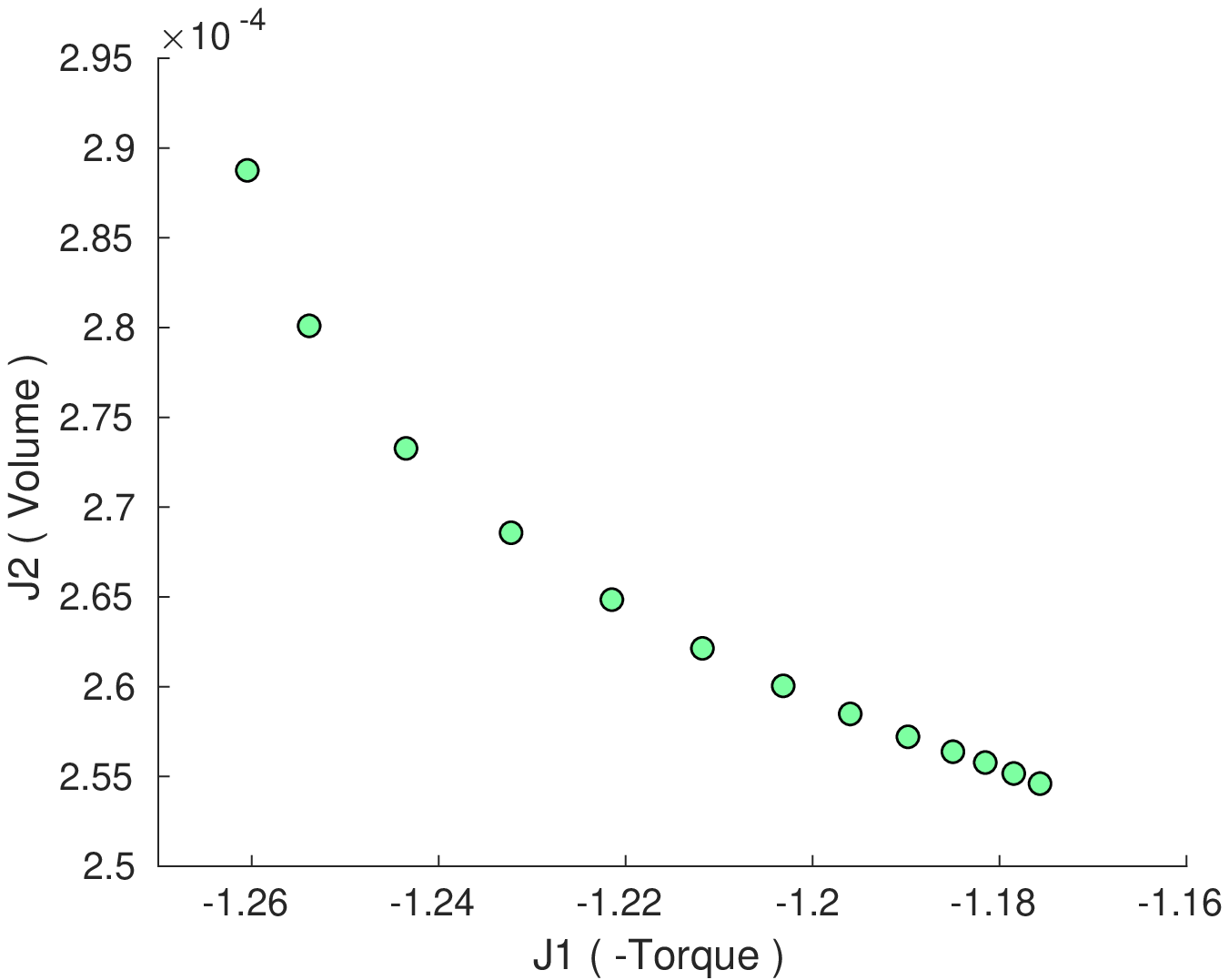}
\end{tabular}
\caption{Left: Values of different designs obtained in the course of gradient based two-objective optimization algorithm for different weights $w$. Right: Zoom on approximated Pareto front. }
\label{fig_pareto}
\end{figure}

\begin{figure}
    \begin{tabular}{ccc}
        \includegraphics[width=.14\textwidth, trim = 500 100 500 100, clip]{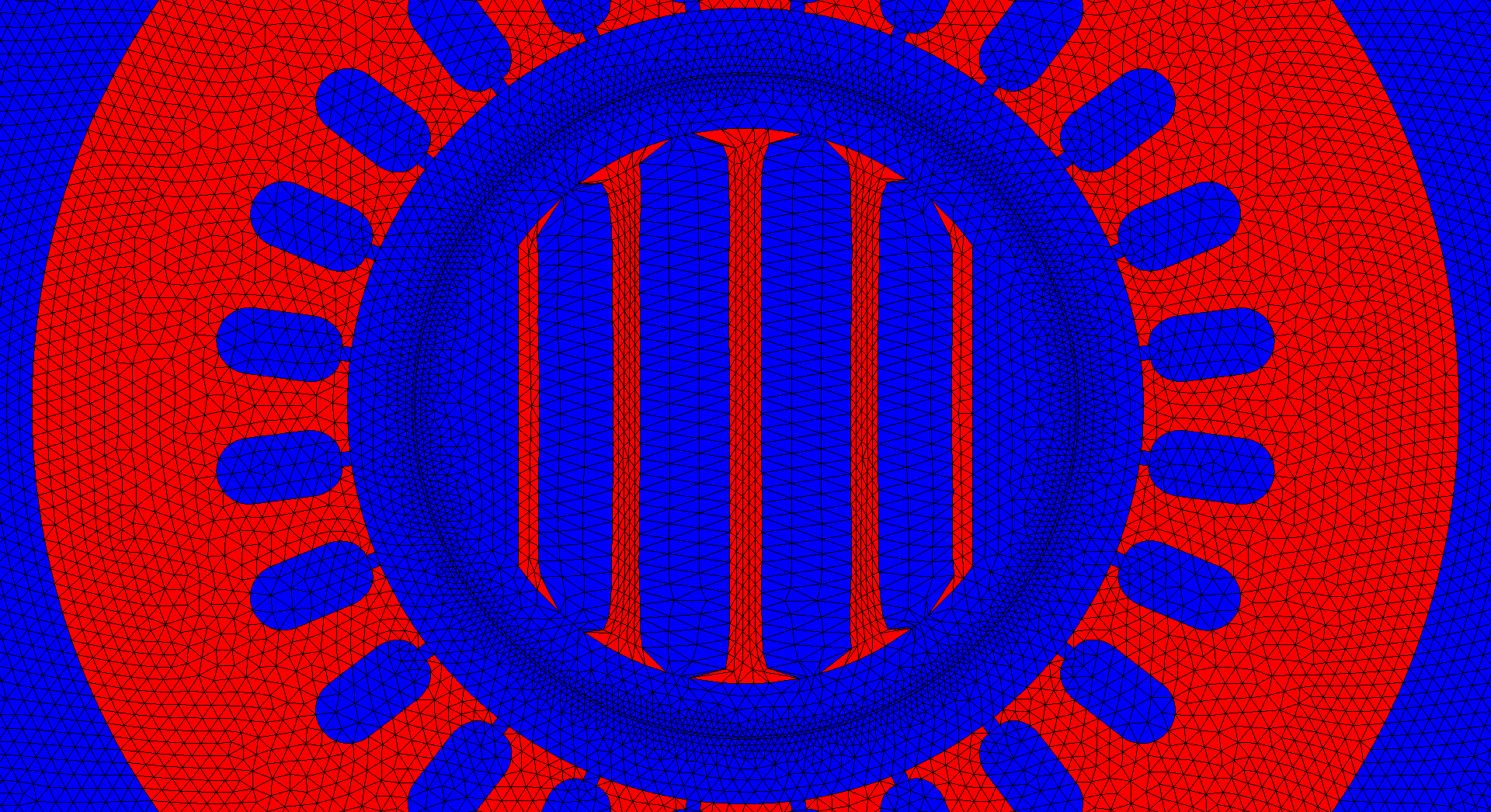} &
        \includegraphics[width=.14\textwidth, trim = 500 100 500 100, clip]{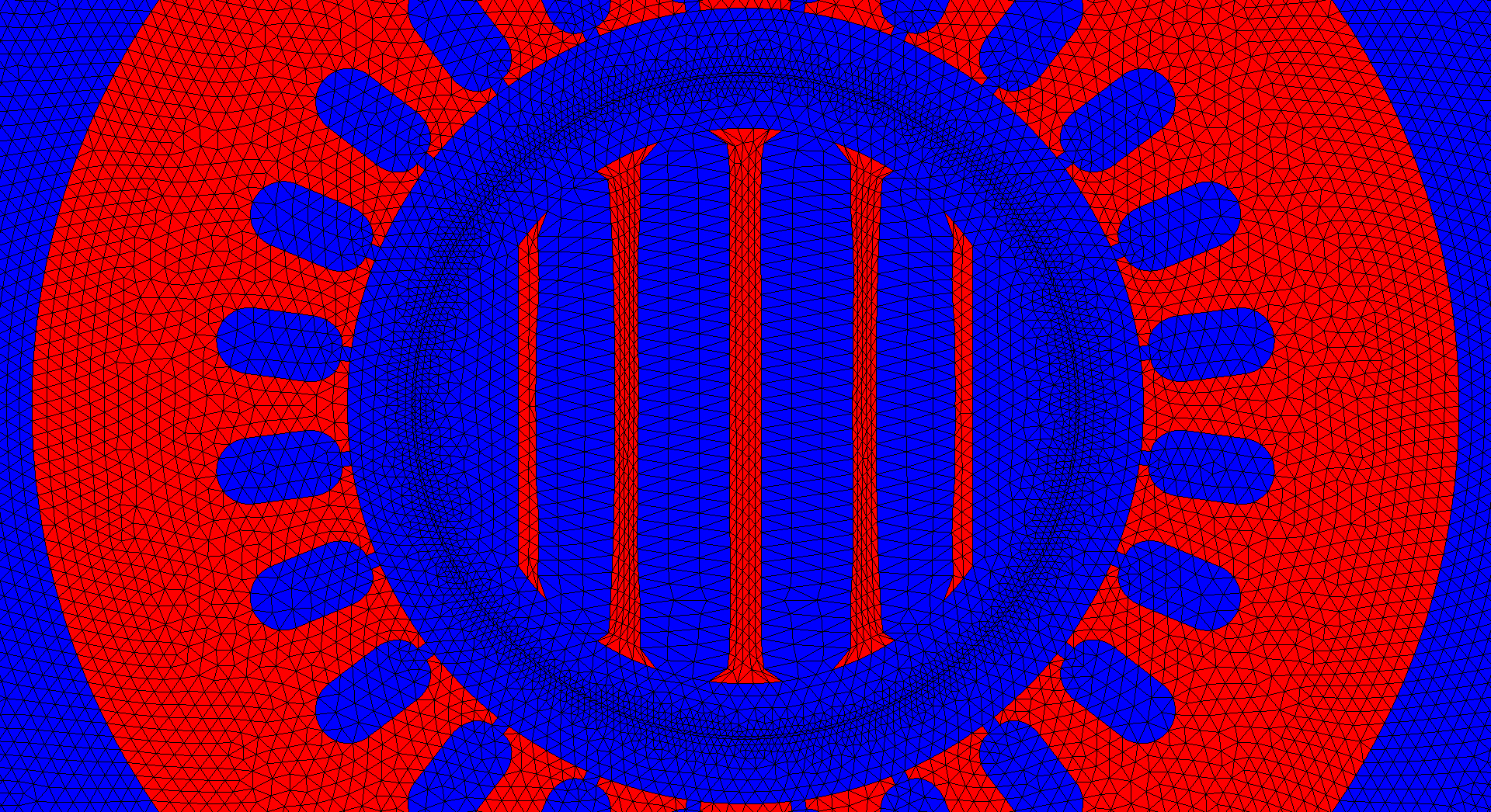} &
        \includegraphics[width=.14\textwidth, trim = 500 100 500 100, clip]{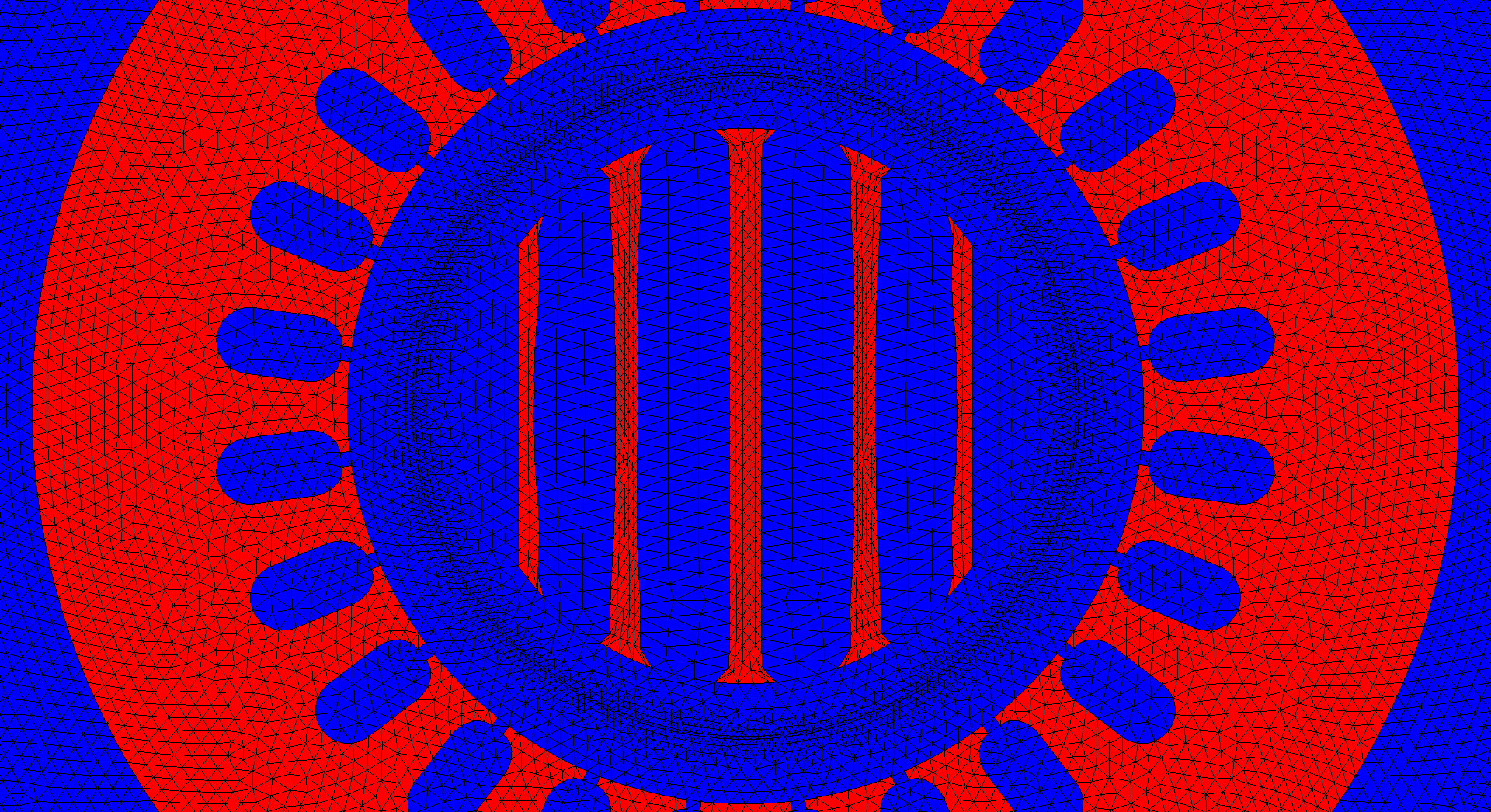} \\
        $w = 0.065$ & $w = 0.035$ &  $w = 0.005$\\
        $T = 1.260$& $T = 1.203$ & $T = 1.176$ \\
        $V = 2.89 \cdot 10^{-4}$ &  $V =2.60 \cdot 10^{-4} $ & $V = 2.55 \cdot 10^{-4} $
    \end{tabular}
    \caption{Three different designs obtained on the approximated Pareto front by using different weights $w$ for $\mathcal J_2$.}
    \label{fig_designsPareto}
\end{figure}

Figure \ref{fig_pareto} (left) shows the results of the bi-objective descent algorithm for minimizing the negative torque and $w$ times the volume, $\mathcal J_1(\Omega) = -\mathcal T(\Omega)$ and $w \mathcal J_2(\Omega) =w \mbox{Vol}(\Omega)$, for different choices of the weighting factor $w$. The right picture of Figure \ref{fig_pareto} depicts a zoom on the obtained Pareto optimal points. The computational effort for obtaining one Pareto optimal design is comparable to the cost of one single-objective optimization run (see Sec.~\ref{sec_numerics_1func}), amounting to a computational time of about two hours on a single core to obtain the depicted Pareto front. The Pareto optimal designs corresponding to three different choices of $w$ can be seen in Figure \ref{fig_designsPareto}.

\section{Conclusion and Outlook} \label{sec_conclusion}
We addressed the problem of finding the optimal shape of the rotor of a synchronous reluctance machine as used in an X-ray tube by means of a gradient-based free-form shape optimization method which is based on the shape derivative. This approach allowed to obtain an optimized shape which exhibits an increase of the torque by 26$\%$ within only several minutes of computation time. The results are confirmed by a geometric parameter optimization in \texttt{JMAG} where the parametrization is motivated by the design obtained by free-form optimization. Moreover, we introduced an extension to the setting of multi-objective shape optimization and showed a way to obtain an approximate Pareto front while significantly decreasing the computation time when compared to evolutionary algorithms.

In this paper we only considered shape optimization approaches which cannot alter the connectivity of the initial design. A next step would be to consider topology optimization methods to additionally allow for changing topologies, in particular in the context of multi-objective optimization. While this was beyond the scope of this paper, it is subject of future work.

\section*{Acknowledgments}
Alessio Cesarano has been funded by the Austrian Science Fund (FWF) project P 32911.

%\vfill

\end{document}